\documentclass{elsart}
\usepackage[dvips]{graphicx}
\usepackage{amsmath}

\begin{document}
\title{Tsallis distribution with complex\\ nonextensivity parameter $q$}
\author{G. Wilk$^{a}$, Z.W\l odarczyk$^{b}$}
\address{$^{a}$National Centre for Nuclear Research,
        Department of Fundamental Research,\\ Ho\.za 69, 00-681
        Warsaw, Poland; e-mail: wilk@fuw.edu.pl\\
         $^{b}$Institute of Physics, Jan Kochanowski University,\\
               \'Swi\c{e}tokrzyska 15; 25-406 Kielce, Poland; e-mail:
               zbigniew.wlodarczyk@ujk.kielce.pl\\
                  }

 {\today}

{\scriptsize Abstract: We discuss a Tsallis distribution with
complex nonextensivity parameter $q$. In this case the usual
distribution is decorated with a log-periodic oscillating factor
(apparently, such oscillations can bee seen in recently measured
transverse momentum distributions in collisions at very high
energies). Complex $q$ also means complex heat capacity which
shall also be briefly discussed.


}

\section{Introduction}
\label{sec:I}

The two parameter Tsallis distribution based on the entropy $S_q$
\cite{Tsallis},
\begin{equation}
f\left( X \right) = C\cdot \left[ 1 + \frac{X}{mT}\right]^{-m}
\label{eq:Tsallis}
\end{equation}
with scale parameter $T$ identified in thermodynamical
applications with the usual temperature (although such
identification cannot be solid \cite{RGC}) and with a real power
index $m=1/(q-1)$ ($q$ being known as the parameter of
nonextensivity in statistical mechanical approaches) are nowadays
well known and applied in a vast variety of situations ($C$ is
normalization constant) \cite{Tsallis}. For $m \rightarrow \infty$
(or $q \rightarrow 1$) this power-like distribution coincides with
the usual exponential distribution $f(X) = C\exp(- X/T)$.
Actually, a Tsallis distribution can be regarded as a
generalization to real power $m$ (or $q$) of such well known
distributions as the Gosset-Student distribution ($X = t^2$, $m =
(\nu+1)/2$ with integer $\nu$, which for $\nu \rightarrow \infty$
becomes a Gaussian distribution and for $\nu = 1$ a Cauchy
distribution).

In this note we investigate the case when $m$ (or $q$) in Eq.
(\ref{eq:Tsallis}) is complex. It turns out that in such a case
the Tsallis distribution retains its main quasi-power like form,
but this form is now decorated with some specific log-periodic
oscillations. In fact, such behavior has been found in many
places, such as earthquakes \cite{EQ}, escape probabilities in
chaotic maps close to crisis \cite{CHM}, biased diffusion of
tracers on random systems \cite{BD}, kinetic and dynamic processes
on random quenched and fractal media \cite{RQFM}, considering
specific heat associated with self-similar \cite{VMdST} or fractal
spectra \cite{TdSMVM}, diffusion-limited-aggregate clusters
\cite{DLM}, growth models \cite{GM}, or stock markets near
financial crashes \cite{FM}, to name only a few examples. However,
in all these cases the main distributions were scale free power
law ones without any scale parameter (here $T$) and without a
constant term governing their $X < mT$ behavior. In the context of
nonextensive statistical mechanics log-periodic oscillations have
first been observed while analyzing the convergence dynamics of
$z$-logistic maps \cite{UT}.

\section{Log-periodic oscillations}
\label{sec:II}

\begin{figure}[h]
\vspace{-3mm}
\begin{center}
\includegraphics[width=8.2cm]{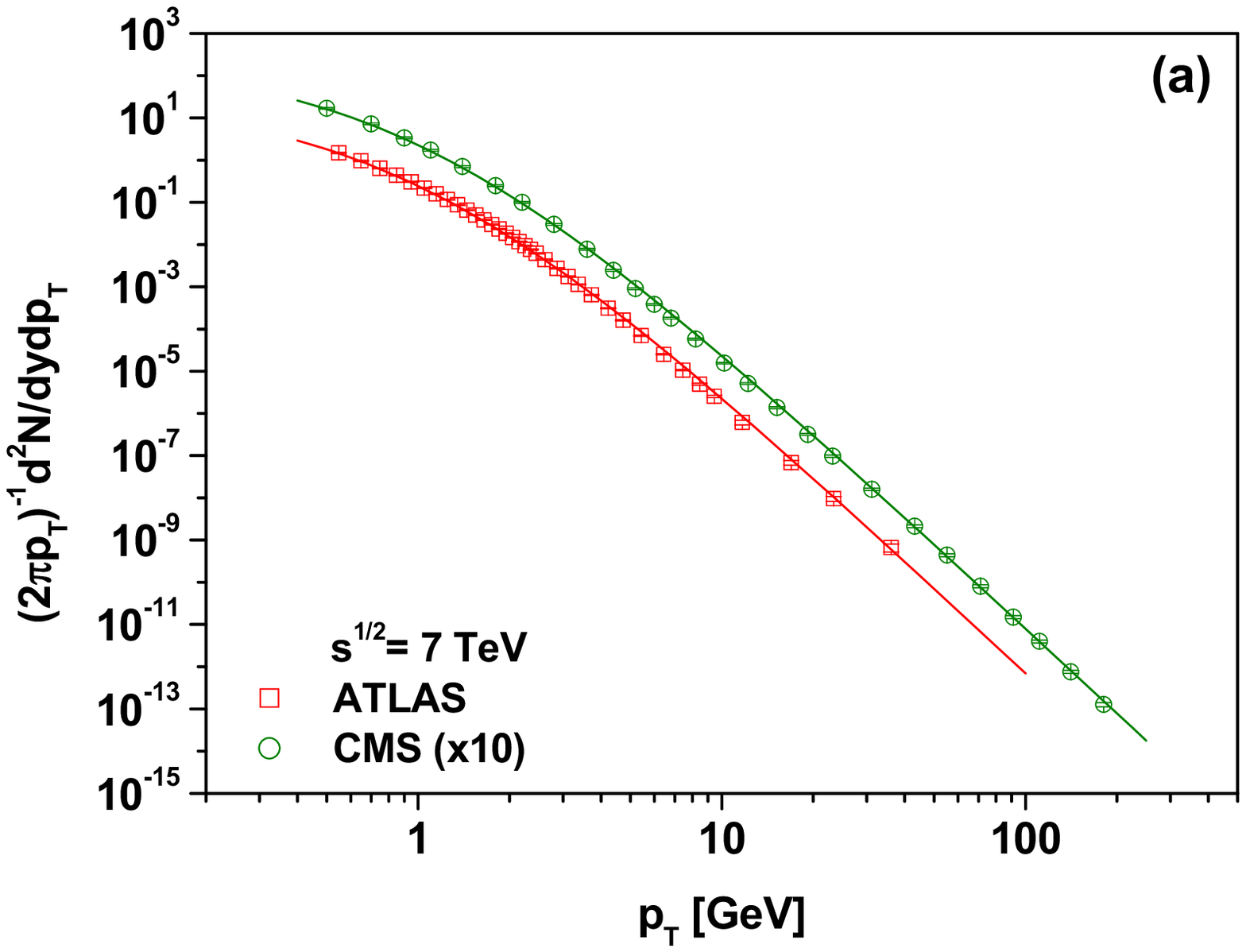}\hfill
\includegraphics[width=8cm]{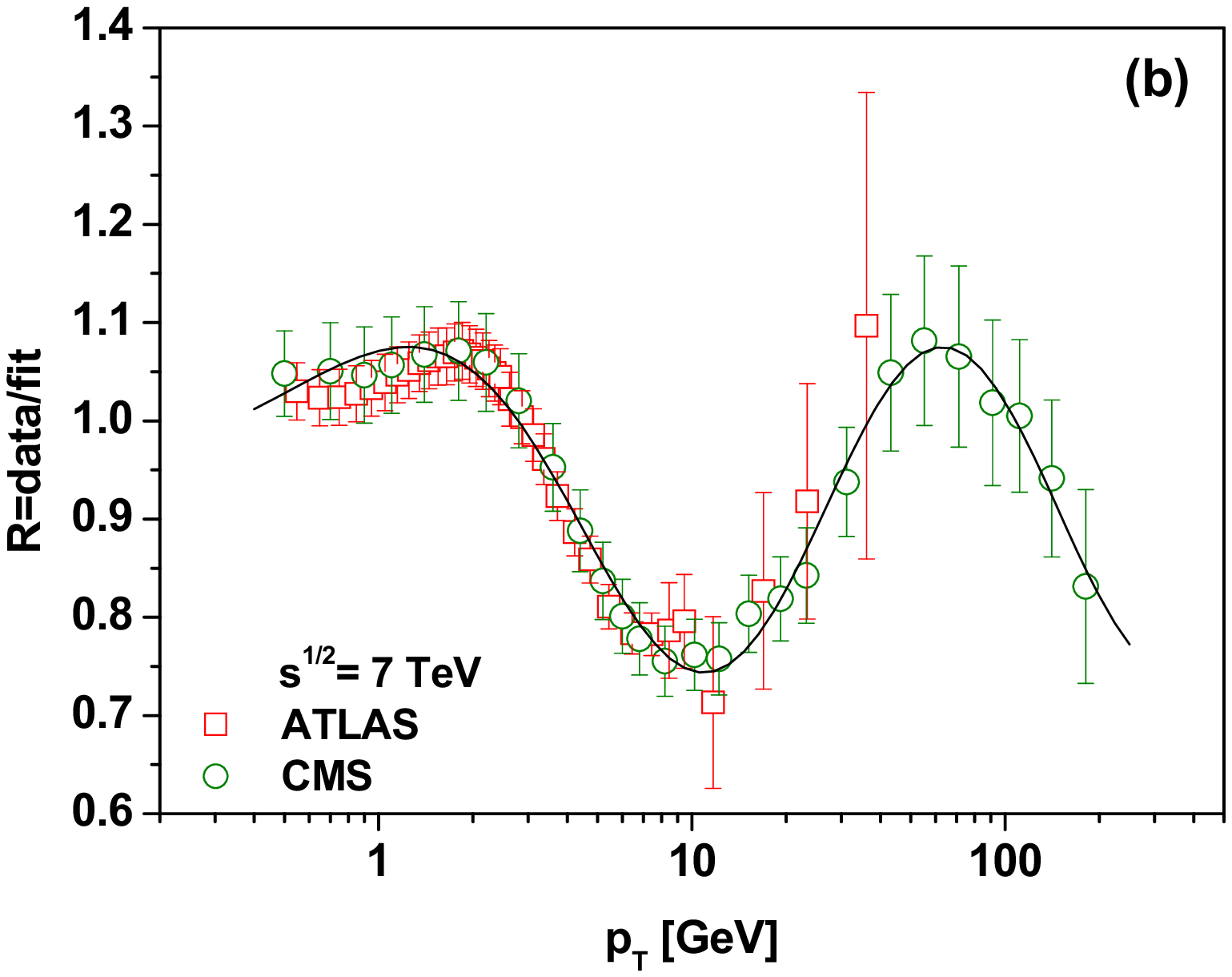}
\vspace{-5mm} \caption{(Color online) $(a)$ Fit to $p_T$ data for
pp collisions at $7$ TeV from CMS \cite{CMS} and ATLAS
\cite{ATLAS} experiments using distribution (1) with parameters
used $T = 0.145$ GeV and $m = 6.7$. Data  points for the CMS
experiment are scaled by a factor of $10$ for better readability.
$(b)$ Fit to $p_T$ dependence of data/fit ratio for results
presented in the left panel $(a)$ using the function $R$ from Eq.
(\ref{eq:Factor}) with: $a = 0.909$, $b = 0.166$, $c = 1.86$, $d =
0.948$ and $f = -1.462$.} \label{Fig1}
\end{center}
\end{figure}

We illustrate our point by an example of recent results obtained
for the highest presently available energies  of $7$ TeV in two
experiments performed at the Large Hadron Collider at CERN, namely
CMS \cite{CMS} and ATLAS \cite{ATLAS}. In Fig. \ref{Fig1}a we show
the observed transverse momentum ($p_T$) distributions for
secondaries produced in proton-proton collisions in these
experiments. These secondaries were produced at midrapidity, i.e.,
for $y = \frac{1}{2}\ln\frac{E + p_L}{E-p_L} \simeq 0$ for which,
for large transverse momentum, $p_T > M$ (where $M$ is the mass of
the particle), one has that, approximately, the energy of
particle, $E = \sqrt{M^2 + p_T^2} \cosh (y) \simeq p_T$, i.e., it
practically coincides with $p_T$ ($p_L = \sqrt{M^2 + p_T^2} \sinh
(y)$ is the longitudinal momentum of the observed particle).
Albeit both fits look pretty good, closer inspection shows that
the ratio of data/fit is not flat. It shows some kind of visible
oscillations, cf. Fig. \ref{Fig1}b. It turns out that these
oscillations cannot be compensated or erased by any reasonable
change of fitting parameters. Instead, to account for them
distributions $f\left( p_T \right)$ from Eq. (\ref{eq:Tsallis})
have to be multiplied by some log-periodic oscillating
factor\footnote{~~Detailed analysis of this phenomenon in the
available high energy experimental data is presented in
\cite{qData}.} (for identification of parameters $a$, $b$, $c$,
$d$ and $f$ used here see Eqs.(\ref{eq:a})-(\ref{eq:f}) below):
\begin{equation}
R(E)= a + b\cos\left[ c\ln(E + d) + f\right]. \label{eq:Factor}
\end{equation}

To explain the origin of such a dressing factor (and tacitly
assuming that it is not an experimental artifact, as it was
observed in both experiments), start from the known observation
that, whereas Boltzmann-Gibbs (BG) distribution,
\begin{equation}
f(E) = \frac{1}{T} \exp\left( - \frac{E}{T}\right), \label{eq:BG}
\end{equation}
comes from the simple equation,
\begin{equation}
\frac{df(E)}{dE} = - \frac{1}{T} f(E), \label{eq:BGde}
\end{equation}
with the scale parameter $T$ being constant, the same equation,
but with variable scale parameter in the form (known as {\it
preferential attachment} in networks
\cite{NETWORKS,qWW1}\footnote{~~~It is worth recalling here that
this very same form, $T(E) = T_0 + (1 - q)E$, also appears in
\cite{WR} within a Fokker-Planck dynamics applied to the
thermalization of quarks in a quark-gluon plasma by a collision
processes.} ),
\begin{equation}
T = T(E) = T_0 + \frac{E}{n},  \label{eq:variableT}
\end{equation}
is now,
\begin{equation}
\frac{df(E)}{dE} = - \frac{1}{T(E)}f(E) = - \frac{1}{T_0 +
E/n}f(E), \label{eq:net}
\end{equation}
and results in the Tsallis distribution
\begin{equation}
f(E) = \frac{n - 1}{nT_0} \left( 1 + \frac{E}{nT_0}\right)^{-n}.
\label{eq:T}
\end{equation}
Now write Eq. (\ref{eq:net}) in finite difference form, namely as
\begin{equation}
f(E + \delta E) = \frac{-n \delta E + nT + E}{nT + E} f(E),
\label{eq:netD}
\end{equation}
(this means, in practical sense, a first-order Taylor expansion
for small $\delta E << E$; from Eq. (\ref{eq:netD}) on, we use $T$
instead of $T_0$) and consider a situation in which $\delta E$
always remains finite (albeit, depending on the value of the new
scale parameter $\alpha$, it can be very small) and equal to
\begin{equation}
\delta E = \alpha nT(E) = \alpha (nT + E).  \label{eq:dEalpha}
\end{equation}
Because one expects that changes $\delta E$ are of the order of
the temperature $T$, the scale parameter must be limited by $1/n$,
i.e., $\alpha < 1/n$. In this case, substituting
(\ref{eq:dEalpha}) into (\ref{eq:netD}), we have,
\begin{equation}
f[E + \alpha(nT + E)] =  (1 - \alpha n)f(E). \label{eq:scaling1}
\end{equation}
Expressing now Eq. (\ref{eq:scaling1}) in a new variable $x$,
\begin{equation}
x = 1 + \frac{E}{nT}, \label{eq:x}
\end{equation}
we recognize that the argument of the function on the left-hand
side of equality (\ref{eq:scaling1}) is $E + \alpha (nT + E) = (1
+ \alpha)xnT - nT$, while the argument of the function on its
right-hand side is $E = xnT - nT$. Notice that, in comparison with
the right-hand side, the variable $x$ on the left-hand side is
multiplied by the additional factor $(1 + \alpha)$. This then
means that, formally, Eq.(\ref{eq:scaling1}), when expressed in
$x$, corresponds to the following scale invariant relation:
\begin{equation}
g[(1 + \alpha )x] = (1 - \alpha n)g(x). \label{eq:scaling}
\end{equation}

Now, it is known \cite{Scaling} that, if for some function $O(x)$,
one finds that $O(\lambda x) = \mu O(x)$ then it is scale
invariant and its form follows a simple power law, $O(x) =
Cx^{-m}$ with $m = - \ln \mu/\ln \lambda$. This relation can be
written as $\mu \lambda^{m} = 1 = e^{i2\pi k}$, where $k$ is an
arbitrary integer. Therefore we have, in general, not a single
power $m$ but rather a whole family of powers, $m_k$, with $m_k =
- \ln \mu/\ln \lambda + i 2\pi k/\ln \lambda$, i.e., it is a
complex number, the imaginary part of which signals a hierarchy of
scales leading to log-periodic oscillations. Coming back to Eq.
(\ref{eq:scaling}), its general, solution is a power law,
\begin{equation}
 g(x) = x^{-m_k}, \label{eq:power}
 \end{equation}
 with exponent $m_k$ depending on $\alpha$ and acquiring an imaginary part,
 \begin{equation}
 m_k = - \frac{\ln ( 1 - \alpha n)}{\ln (1 +
 \alpha)} + ik \frac{2\pi}{\ln(1 + \alpha)}. \label{eq:solution}
\end{equation}
The special case of $k=0$, i.e., the usual real power law solution
with $m_0$ corresponding to fully continuous scale
invariance\footnote{~~~~~~In this case power law exponent $m_0$
still depends on $\alpha$ and increases with it roughly as $m_0
\simeq n + \frac{n}{2} (n+1)\alpha + \frac{n}{12}\left(4n^2 + 3n
-1\right)\alpha^2 + \frac{n}{24}\left( 6n^3 + 4n^2 - n
+1\right)\alpha^3 + \dots$. Notice also that $ \alpha < 1/n$. },
recovers in the limit $\alpha \rightarrow 0$ the power $n$ in the
usual Tsallis distribution. In general one has
\begin{equation}
g(x) = \sum_{k=0}w_k\cdot {\rm Re}\left( x^{-m_k}\right) = x^{-
{\rm Re}\left( m_k\right)}\sum_{k=0}w_k\cdot \cos\left[ {\rm
Im}\left(m_k\right) \ln(x) \right]. \label{eq:fin}
\end{equation}
One therefore obtains a Tsallis distribution decorated by a
weighted sum of log-oscillating factors (where $x$ is given by Eq.
(\ref{eq:x})). Because usually in practice we do not {\it a
priori} know the details of the dynamics of processes under
consideration (i.e., we do not known the weights $w_k$), for
fitting purposes one usually uses only $k=0$ and $k=1$. In this
case one has, approximately,
\begin{equation}
g(E) \simeq \left( 1 + \frac{E}{nT}\right)^{-m_0}\left\{ w_0 +
w_1\cos\left[ \frac{2\pi}{\ln (1 + \alpha)} \ln \left( 1 +
\frac{E}{nT}\right)\right]\right\} \label{eq:approx}
\end{equation}
and reproduces the general form of a dressing factor given by Eq.
(\ref{eq:Factor}) and often used in the literature \cite{Scaling}.
In this approximation the parameters $a$, $b$, $c$, $d$ and $f$
from Eq. (\ref{eq:Factor}) get the following meaning:
\begin{eqnarray}
\frac{a}{b} \,&=&\, \frac{w_0}{w_1}; \label{eq:a}\\
c\, &=&\, \frac{2\pi}{\ln (1 + \alpha)}; \label{eq:c}\\
d\, &=&\, n T; \label{eq:d}\\
f\, &=&\, - \frac{2\pi}{\ln(1 + \alpha)} \ln(nT). \label{eq:f}
\end{eqnarray}

 However, this is not the most general result possible. Notice
that in our derivation presented by
Eqs.(\ref{eq:dEalpha})-(\ref{eq:scaling})) we only accounted for a
single step evolution, whereas in reality we can have a whole
hierarchy of evolutions. Then one has that
\begin{equation}
E_i = E_{i-1} + \alpha_{i-1} \left( nT + E_{i-1} \right),
\label{eq:cascade}
\end{equation}
each with its own scale parameter $\alpha_i$. In the simplest
situation, neglecting any fluctuations of consecutive scaling
parameters, i.e., assuming that all $\alpha_i = \alpha$, after
$\kappa$ steps one has that
\begin{equation}
nT + E_{\kappa} = ( 1 + \alpha)^{\kappa} \left( nT + E_0 \right).
\label{eq:steps}
\end{equation}
This means then that Eq. (\ref{eq:scaling}) should be replaced by
a new scale invariant equation:
\begin{equation}
g\left[ (1 + \alpha)^{\kappa} x\right] = ( 1 - \alpha n)^{\kappa}
g(x). \label{eq:scallingK}
\end{equation}
Whereas this equation does not change the slope parameter $m_0$,
it significantly influences the frequency of oscillations which
are now $\kappa$ times smaller,
\begin{equation}
c = \frac{2 \pi}{\kappa \ln (1 + \alpha)} \label{eq:kappa}
\end{equation}
(in Eq.(\ref{eq:scallingK}) $\lambda = (1 + \alpha)^{\kappa}$ and
$\mu = (1 - \alpha n)^{\kappa}$; the slope parameter $m_0 = - \ln
\mu/\ln \lambda$ is independent of $\kappa$, whereas the frequency
of oscillations, $2\pi/\ln \lambda$, decreases with $\kappa$ as
$1/\kappa$). For more complex behavior of intermediate scale
parameters $\alpha_i$ one gets more complicated expressions (we
shall not discuss this here).

\section{Other consequences of complex nonextensivity parameter}
\label{sec:III}

There are other consequences of allowing the parameter $m$ to be
complex. Namely, the complex power exponent in the Tsallis
distribution, $m = m' + i\cdot m''$, also means a complex
nonextensivity parameter $q$,
\begin{equation}
q = 1 + \frac{1}{m} = q' + i\cdot q'' \label{eq:c_q1}
\end{equation}
where \begin{equation} q' = 1 + \frac{m'}{|m|^2};\qquad q'' = -
\frac{m''}{|m|^2}. \label{eq:c_q2}
\end{equation}
Now, the complex nonextensive parameter $q$ has some profound
consequences. This is because, as shown in \cite{BiroC} (and also
in \cite{Campisi,HC_old,qWW1,qWW}), the nonextensivity parameter
$q$ can be treated as a measure of the thermal bath heat capacity
$C$ with
\begin{equation}
C = \frac{1}{q - 1}.\label{eq:HC}
\end{equation}
It means therefore that, in general, the heat capacity becomes
complex as well. As a matter of fact, such complex (frequency
dependent) heat capacities (or generalized calorimetric
susceptibilities) are known in the literature \cite{HC} under the
form
\begin{equation}
C = C' -i C'' = C_{\infty} + \frac{C_0 - C_{\infty}}{1 +  (\omega
\tau)^2}(1 - i \omega \tau).  \label{eq:hc}
\end{equation}
Here $C_{\infty}$ is the heat capacity related to the infinitely
fast degrees of freedom of the system as compared to the frequency
$\omega$, and $C_0$ is the total contribution at equilibrium (the
frequency is set to zero) of the degrees of freedom, fast and
slow, of the sample. The time constant $\tau$ is the kinetic
relaxation time constant of a certain internal degree of freedom.

These complex heat capacities are known as dynamic heat capacities
and are intensively explored from both experimental and
theoretical perspectives. It is expected that dynamic calorimetry
can provide an insight into the energy landscape dynamics, cf.,
for example, \cite{G2007,GR2007,ND,S}. Usually one associates the
imaginary part of linear susceptibility with the absorption of
energy by the sample from the applied field.

In the case of temperature fluctuations  $ \delta T(t)$ the
deviation of the energy from its equilibrium value  $\delta U(t)$
is, for a certain linear operator $\hat{C}(t)$, some linear
function of the corresponding variation of the temperature,
\begin{equation}
\delta U(t) = \hat{C} \delta T(t). \label{eq:hatC}
\end{equation}
If the temperature of the reservoir changes infinitely slowly in
time, then the system can keep up with any changes in the
reservoir and its susceptibility is just the specific heat of the
system $C_V$. However, in general, the behavior of the system is
described by a generalized susceptibility $C_V(\omega)$, which can
be called {\it the complex and $\omega$-dependent} heat capacity
of the system\footnote{~~~~~~~~The change in the energy of a
system in the field of the thermal force can be represented by
$\delta U(t) = \int L\left(t'\right)\delta T\left(t -
t'\right)dt'$ where $L\left( t'\right)$ is the response function
of the system describing its relaxation properties given by
$\Phi(t) = \int_t^{\infty}L\left( t'\right)dt'$. Taking the
Fourier transform one gets $\delta U(\omega) = C_V(\omega)\delta
T(\omega)$ where $C_V(\omega) = \int L\left( t'\right)e^{i\omega
t'} dt'$ is the generalized susceptibility of the system and is
called the complex heat capacity. In practice, the frequency
dependent heat capacity is a linear susceptibility describing the
response of the system to the small thermal perturbation
(occurring on the time scale $1/\omega$)  that takes the system
slightly away from the equilibrium .}.

A complex $C_V(\omega)$  means that $\delta U$  and $\delta T$ are
shifted in phase and that the entropy production in the system
differs from zero \cite{S}. The corresponding
fluctuation-dissipation theorem for the frequency dependent heat
capacity was established in \cite{ND}. According to this result,
the frequency-dependent heat capacity may be expressed within the
linear response approximation as a linear susceptibility
describing the response of the system to arbitrarily small
temperature perturbations away from equilibrium,
\begin{equation}
C_V(\omega) = \frac{1}{T_0^2} \left( \langle U^2\rangle_0 - i
\omega \int_0^{\infty} dt e^{-i\omega t} \langle U(0) U(t)\rangle
\right) \label{eq:C_V}
\end{equation}
(the $\omega$ denotes frequency with which temperature field is
varying with time).

The above results for heat capacity can now be used to a new
phenomenological interpretation of the complex $q$ parameter
discussed before. Namely, one can argue that (we denote now $T_0 =
\langle T\rangle$)
\begin{equation}
q - 1 = \frac{Var(T)}{\langle T\rangle^2} - i \frac{S(T)}{\langle
T\rangle^2} \label{eq:qCq}
\end{equation}
where
\begin{equation}
S(T) = \omega \int \langle Cov[ T(0),T(t) ]\rangle e^{-i\omega t}
dt \label{eq:ST}
\end{equation}
is the spectral density of temperature fluctuations (i.e., the
Fourier transform of the covariance function averaging over the
nonequilibrium density matrix)\footnote{~~~~~~~~~~~~We would like
to stress at this point that, in a sense, Eq. (\ref{eq:qCq}) can
be regarded as a generalization of our old proposition for
interpreting $q$ as a measure of nonstatistical intrinsic
fluctuations in the system \cite{WWq} (which corresponds to the
real part of (\ref{eq:qCq})) by adding the effect of spectral
density of such fluctuations (via the imaginary part of
(\ref{eq:qCq})). Notice that (\ref{eq:qCq}) follows from
(\ref{eq:HC}) and the relation $U=C_V T$, allowing to write
(\ref{eq:C_V}) in the form of (\ref{eq:qCq}). }.

\section{Conclusions}
\label{sec:IV}

To summarize: Log-periodic structures in the data indicate that
the system and/or the underlying physical mechanisms have
characteristic scale invariance behavior. This is interesting as
it provides important constraints on the underlying physics. The
presence of log-periodic features signals the existence of
important physical structures hidden in the fully scale invariant
description. It is important to recognize that Eq. (\ref{eq:net})
represents an averaging over highly 'non-smooth' processes and, in
its present form, suggests rather smooth behavior. In reality,
there is a discrete time evolution for the number of steps. To
account for this fact, one replaces a differential  Eq.
(\ref{eq:BGde}) by a difference quotient and expresses $dt$ as a
discrete step approximation given by Eq. (\ref{eq:dEalpha}) with
parameter $\alpha$ being a characteristic scale ratio. It can also
be shown that discrete scale invariance and its associated complex
exponents can appear spontaneously, without a pre-existing
hierarchical structure.

\section*{Acknowledgements}

This research  was supported in part by the National Science
Center (NCN) under contract DEC-2013/09/B/ST2/02897. We would like
to warmly thank Dr Eryk Infeld for reading this manuscript.


\begin{thebibliography}{99}



\bibitem{Tsallis} C.~Tsallis, J.\ Statist.\ Phys.\ {\bf 52} (1988) 479 (1988);
                  Eur.\ Phys.\ J.\ A {\bf 40} (2009) 257
                  and {\it Introduction to Nonextensive Statistical Mechanics}
                  (Springer, 2009). For an updated bibliography on this subject,
                  see http://tsallis.cat.cbpf.br/biblio. htm.

\bibitem{RGC}     C.~Tsallis, Physic\ A{\bf 221} (1995) 277;
                  L.~A.~Rios, R.~M.~O.~Galv\~{a}o and L.~Cirto, Phys.\ Plasmas
                  {\bf 19} (20012) 034701.


\bibitem{EQ}      Y.~Huang, H.~Saleur, C.~Sammis  and D.~Sornette,
                  Europhys.\ Lett.\ {\bf 41} (1998) 43;
                  H.~Saleur, C.~G.~Sammis and  D.~Sornette, J.\ Geogphys.\ Res.\
                  {\bf 101} (1996) 17661.

\bibitem{CHM}     A.~Krawiecki, K.~Kacperski, S.~Matyjaskiewicz  and J.~A.~Holyst,
                  Chaos,\ Solitons,\ Fractals {\bf 18} (2003) 89.

\bibitem{BD}      J.~Bernasconi and  W.~R.~ Schneider, J.\ Stat.\ Phys.\ {\bf 30} (1983)
                  355; D.~Stauffer  and  D.~Sornette , Physica\ A {\bf 252} (1998) 271;
                  D.~Stauffer, Physica\ A {\bf 266} (1999) 35.

\bibitem{RQFM}    B.~Kutnjak-Urbanc, S.~Zapperi, S.~Milosevic and H.~E.~Stanley, Phys.\
                  Rev. E {\bf 54} (1996) 272; R.~F~.~S.~Andrade, Phys.\ Rev.\ E {\bf 61}
                  (2000) 7196;  M.~A.~Bab, G.~Fabricius and  E.~V.~Albano, Phys.\ Rev.\ E
                  {\bf 71} (2005) 36139; H.~Saleur  and  D.~Sornette, J.\ Phys.\ I {\bf 6}
                  (1996) 327.

\bibitem{VMdST}   R.~O.~Vallejos, R.~S.~Mendes, L.~R.~da Silva and C~. Tsallis. Phys.\ Rev.\
                  E {\bf 58} (1998)1346.

\bibitem{TdSMVM}  C.~Tsallis, L.~R.~da Silva, R.~S.~Mendes, R.~O.~Vallejos and A.~M.~ Mariz,
                  Phys.\ Rev.\ E {\bf 56} (1997) R4922.

\bibitem{DLM}     D.~Sornette, A.~Johansen, A.~Arneodo, J.-F.~Muzy and H.~Saleur,
                  Phys.\ Rev.\ Lett.\ {\bf 76} (1996) 251.

\bibitem{GM}      Y.~Huang, G.~Ouillon,  H.~Saleur  and D.~Sornette, Phys.\ Rev.\ E {\bf 55}
                 (1997) 6433.

\bibitem{FM}      D.~Sornette, A.~Johansen  and J.-P.~ Bouchaud , J.\ Phys.\ I {\bf 6}
                 (1996) 167;  N.~Vanderwalle, Ph.~Boveroux, A.~Minguet  and M.~Ausloos,
                  Physica\ A {\bf 255} (1998) 201;   N.~Vanderwalle  and M.~Ausloos,
                  Eur.\ J.\ Phys.\ B {\bf 4} (1998) 139;  N.~Vanderwalle, M.~ Ausloos,
                  Ph.~Boveroux  and A.~Minguet, Eur.\ J.\ Phys.\ B {\bf 9} (1999)
                  355; J.~H.~Wosnitza and J.~Leker, Physica\ A {\bf 401} 228.

\bibitem{UT}      F.~A.~B.~F. de Moura, U.~Tirnakli and M.~L.~Lyra,  Phys.\ Rev.\ E {\bf 62}
                  (2000) 6361.

\bibitem{CMS}     V.~Khachatryan $et~al.$ (CMS Collaboration), JHEP\ {\bf 02} (2010) 041
                  and JHEP\ {\bf 08} (2011) 086; Phys.\ Rev.\ Lett.\  {\bf 105} (2010) 022002.

\bibitem{ATLAS}   G.~Aad $et~al.$ (ATLAS Collaboration), New\ J.\ Phys.\ {\bf 3} (2011) 053033.

\bibitem{qData} G~ Wilk and Z.~W\l odarczyk, {\it Log-periodic oscillations of transverse
                momentum distributions}, arXiv:1403.3508v2.

\bibitem{NETWORKS}  G.~Wilk and Z.~W\l odarczyk, Acta\ Phys.\ Polon.\ B {\bf 35} (2004) 871
                    and B {\bf 36} (2005) 2513.

\bibitem{qWW1}  G.~Wilk and Z.~W\l odarczyk, Eur.\ Phys.\ J.\ A {\bf 48} (2012) 161.

\bibitem{WR} D.~B.~Walton and J.~Rafelski, Phys.\ Rev.\ Lett.\  {\bf 84} (2000) 31–34.

\bibitem{Scaling} D.~ Sornette, Phys.\ Rep.\ {\bf 297} (1998) 239.

\bibitem{BiroC} T.~S.~ Bir\'o, G.~G.~Barnaf\"oldi and P.~V\'an, Eur.\ Phys.\ J.\ A {\bf 49}
                (2013) 110.

\bibitem{Campisi} M.~Campisi, Phys.\ Lett.\ A {\bf 366} (2007) 335.

\bibitem{HC_old} A~.R.~Plastino and A.~Plastino, Phys.\ Lett.\ A {\bf 193} (1994) 140;
                 M.~P.~Almeida, Physica\ A {\bf 300} (2001) 424.

\bibitem{qWW} G.~Wilk and Z.~W\l odarczyk, Cent.\ Eur.\ J.\ Phys.\ {\bf 10} (2012) 568.

\bibitem{HC} J.~E.~K.~Schawe, Thermochim.\ Acta\ {\bf 260} (1995) 1; J~.-L.~Garden,
             Thermochimica\ Acta\ {\bf 460} (2007) 85.

\bibitem{G2007} J.~L.~Garden, Thermochim.\ Acta {\bf 452} (2007) 85.

\bibitem{GR2007} J.~L.~Garden and J.~Richard, Thermochim.\ Acta {\bf 461} (2007) 57.

\bibitem{ND} J.~K.~Nielsen and C.~Dyre, Phys.\ Rev.\ B {\bf 54}
             (1996) 15754.

\bibitem{S} G.~I.~Salistra, Sov. Phys. JETP {\bf 26} (1968) 173.

\bibitem{WWq}  G.~Wilk and Z.~W\l odarczyk, Phys.\ Rev.\ Lett.
               {\bf 84} (2000) 2770; T.~S.~ Bir\'o and
               A.~Jakov\'ac, Phys.\ Rev.\ Lett.\ {\bf 94} (2005)
               132302.

\end{thebibliography}
\end{document}